\documentclass[preprint,showpacs]{revtex4} 
\usepackage{graphicx,amsmath}   
\begin{document}  
\title{Pre-avalanche instabilities in a granular pile}      
\author{Lydie Staron$^1$, Jean-Pierre Vilotte$^1$, and Farhang Radjai$^2$}   
\affiliation{$^1$ IPGP, 4 Place Jussieu, F-75252 Paris cedex 05, France. \\   
$^2$ LMGC, CNRS-Universit\'e Montpellier II,   
Place Eugène Bataillon.  
F-34095 Montpellier cedex, France.}

\begin{abstract}  
 
We investigate numerically the transition between static equilibrium and dynamic surface flow of a $2D$ cohesionless granular system driven by a continuous gravity loading. This transition is characterized by intermittent local dynamic rearrangements and can be described by an order parameter defined as the density of critical contacts, e.g. contacts where the friction is fully mobilized. Analysis of the spatial correlations of critical contacts shows the occurence of ``fluidized'' clusters which exhibit a power-law divergence in size at the approach of the stability limit. The results are compatible with recent models that describe the granular system during the static/dynamic transition as a multi-phase system.   

\end{abstract}  

 
\maketitle  

\indent Avalanches and debris flows are of special interest both for
 industrial and natural processes. Although many advances
have been made in the physical understanding of granular flow 
both from a microscopic and a continuum point of view 
 (see for example~\cite{savage89,bouchaud95a,douady99,campbell90,pouliquen99,rajchenbach02}), the  mechanisms leading to
 the transition of a granular material 
from a static equilibrium to a dynamic state are still unclear, and
difficult to analyze experimentally. 
Recent hydrodynamic model based on phase transition theory
describes the granular material as a two-phase material 
with a ''solid'' (static) and a ``liquid'' (flowing) phase~\cite{aranson01a}.
This transition, also advocated in others conditions~\cite{mueth00,metcalfe02}, is important for the understanding of avalanche
instability in the context of risk assessment. \\
\indent The purpose of this paper is to investigate numerically 
this static to dynamic transition in a minimal model configuration, 
e.g. a 2D cohesionless granular medium driven by continuously 
loading body force, here a continuous tilt
under gravity. We find that the evolution is chacterized by intermittent 
local instabilities and can be described by the fraction 
of contacts where friction force is completely mobilized. A more detailed 
description involves ``fluidized'' areas which percolate 
when the system reaches the stability limit. 
These results would be in favor of a description in terms 
of a multiphase system undergoing a phase 
transition.\\
%
\indent The contact friction is described by the classical 
Coulomb's law. Relative slip between two particles in
contact can occur only when the friction force is fully activitated, 
i.e $f_t = \pm \mu f_n$, where $f_t$ and $f_n$ are respectively
 the tangential and normal forces at the contact 
and $\mu$ is the contact friction coefficient. This defines the
Coulomb threshold. Otherwise, no slip can occur and 
$f_t \in [-\mu f_n, \mu f_n]$.  
The numerical simulations were performed using the Contact Dynamics 
method~\cite{jean92,moreau94}, based on a fully implicit resolution of
the contact forces. This allows for an accurate 
determination of the ratio $ f_t/f_n$ independently of
any numerical regularization parameter. In all the simulations, the
 contact friction coefficient is $\mu=0.5$ and collisions are perfectly inelastic.\\
%
\indent The initial configuration of the pile is 
generated by random deposition of $N_p=4000$ 
disks in a rectangular box; the disks have a uniform distribution 
of diameters within the range $[D_{min},D_{max}]$. We used the 
ratio $D_{max}/D_{min}=1.5$, but we checked that 
for a polydispersity as large as $D_{max}/D_{min}=10$ the results 
remain essentially unchanged. The granular samples prepared by random 
deposition have a rectangular shape with a nearly 
flat surface, a thickness of $30D$ and a width of $120D$, 
where $D$ is the mean disk diameter. The initial 
coordination number, {\it i.e.} 
the mean number of contacts of a particle,  
is $z \simeq 3.6$.
 It shows only weak fluctuations (up to $5\%$) 
in the course of tilting. The results  
presented below were obtained from $15$ independent runs  
in which the random processing of grain sizes 
is the only source of noise.\\
%
\indent The granular bed is slowly tilted with a constant 
rotation rate ($0.001^\circ$ per time step). 
The slope increases from $\theta=0^\circ$ 
to the maximum angle of repose $\theta_c \simeq 20^\circ$
at which a surface avalanche occurs.  Within the picture of 
ideal Coulomb's material, $\theta_c$ is
related to the internal coefficient of 
friction of the pile $\mu_{eff}$ 
through the relation $\mu_{eff} = tan(\theta_c)\simeq 0.35$~\cite{roux2001}.
However, the evolution of the pile towards $\theta_c$ is
not monotonous. Indeed, we observe the occurence of
local instabilities owing to the mobilization of friction 
between particles. Initially $f_t$ at the contacts
is only partially activated, namely
 $|f_t|<\mu f_n$.
But upon tilting, a number of contacts reach the 
Coulomb threshold, {\it i.e.} $|f_t|=\mu f_n$. 
These contacts can not sustain further shear force increment
and can lead eventually to a slip instability or the 
disappearance of the contact.
 We call them {\em critical contacts}. 
We describe the evolution of the pile in terms  
 of the fraction $\nu$ of critical contacts  
over a volume $V$ and as a function of $\theta$:
\begin{equation}
\nu(\theta, V)=\left<\frac{N_c}{N}\right>_{V},
\label{eqn1}
\end{equation}  
where $N_c$ and $N$ are respectively the number of critical  
contacts and the total number of contacts in $V$ at slope angle $\theta$. 
Since the density of contacts remains almost constant 
due to close-packing ($N \propto V$), $\nu$ also represents 
the density of critical contacts. It characterizes the plastic
state of the pile.\\    
%
%
\indent The evolution of $\nu(\theta, V_{pile})$ is 
reproducible from run 
to run even though there are rapid fluctuations of $\nu$ 
within the system~(see Fig~\ref{fig1}(a)). 
 When averaging the $\nu$ evolution over 15
 independant realizations (inset graph in Fig~\ref{fig1}(a)), we observe a
 regular increase of $\nu(\theta)$ from an
intial value $\nu = 0$ to a maximum value $\nu \simeq 0.08$ at
 the maximum angle of repose $\theta_c$. This indicates that a
partial plastification occurs well before 
the stability limit of the pile. Such a transition is
however characterized by rapid fluctuations of $\nu$. 
They are the signature of intermittent 
instabilities during which loss of critical contacts 
occurs through local dynamical rearrangements, as can be
seen in the fluctuations of the mean kinetic energy in Fig~\ref{fig1}(b). 
This suggests that even though the density
of critical contacts increases in the mean during the 
mobilization stage of the system, the population of
critical contacts is renewed by these intermittent instabilities
and a single critical contact is
only of short-life time due to its metastable state. 
During the intermittent evolution, the system can 
explore extreme $\nu$-states as seen in Fig~\ref{fig1}(a). When 
averaging over small $\theta$ size windows the 
extremal states at each $\theta$ and for all 
independant realizations, their evolution
 exhibits a well defined limit envelope (see Fig~\ref{fig1}(c)). It shows  
an exponential convergence towards an asymptotic
limit characterized by a critical state $\nu_c \simeq 0.08$, 
corresponding to the stability limit. 
The granular system can therefore be described in 
terms of an order parameter $\Phi =\nu/\nu_c$ which 
 varies from zero to one and characterizes 
the partially fluidized transition.\\
\indent Owing to the intrinsic geometrical and stress fluctuations of 
each individual realization,
each pile can explore 
rare extremal states beyond the critical limit $\nu_c$
in the course of its evolution. These 
states are metastable and followed eventually by local 
dynamical rearrangements, the size of which can be 
characterized by the fraction $\Delta N/N$ of contacts lost in 
that event. When analysing the size of the 
local rearrangements as a function of the 
$\nu$-state explored by the system prior to the event (see Fig~\ref{fig2}), 
we found that the pile evolution is
characterized by frequent weak rearrangements, less than 1$\%$, 
that corresponds to $\nu$-states of the
system below $\nu_c$, and rare and strong rearrangements 
corresponding to extremal states $\nu> \nu_c$. 
The intermittent rearrangements 
during the mobilization occurs both in the bulk and 
close to the free surface.  \\
%
%
The evolution of $\nu$ may be also interpreted in 
terms of the mean separation distance $\lambda(\theta)$
between two critical contacts: 
\begin{equation}
\lambda(\theta) =\sqrt{\frac{2}{z(\theta)\nu(\theta)}}D
\label{eqn2}
\end{equation}
where $z(\theta)$ is the coordination number.
During the mobilization, $\lambda$ decreases from $+\infty$ 
when $\nu=0$ to $\lambda_c= 2.5D$  when 
$\nu=\nu_c$. For $\lambda <\lambda_c$, {\it i.e.} $\nu>\nu_c$, 
the packing undergoes large-scale instabilities due to 
spatial correlations among critical contacts (see Fig.~\ref{fig2}).\\ 
%
\indent To investigate the distribution 
of critical contacts, we introduce
the probability density function (pdf) $P(\nu)$ of 
 finding a local state $\nu$ within a contact neighbourghood.
 The size of this neighbourghood  must not be 
smaller than the minimum length
separating two critical contacts, of the order of $D$, nor
larger than the typical size of spatial heterogeneities of $\nu$
 (to be analyzed later).  
 In particular, we introduce the pdf $P_c(\nu)$ at critical 
contacts and the pdf $P_{nc}(\nu)$ at non-critical contacts.
The distributions $P_c$ and $P_{nc}$ time-averaged over all 
the loading phase from $\theta = 0$ to $\theta_c$
 are displayed in Fig~\ref{fig3}.   
The range of local states $\nu$
is quite wide for both distributions: $P_c$ has a well-defined 
peak at $\nu \simeq \nu_c$, whereas $P_{nc}$ is a 
decreasing function with a high peak at $\nu=0$.   
This indicates that critical contacts tend to appear preferentially 
in the vicinity of other critical contacts, and are 
strongly correlated spatially.
It is worth to note here that the most probable local state is found
to be the critical state $\nu_c$ that characterizes the whole pile at the 
stability limit. \\
%
%
\indent We study the spatial correlations
by defining the mean state $\bar{\nu}$ in a neighbourghood of 
critical contacts as a function of the neighbourghood's
size $r$ and of the inclination angle $\theta$ of the pile.
In practice, we explore $r\in[1D, 20D]$.
Denoting $\nu_i$ the local state in the vicinity  
of a contact $i$, $\bar{\nu}(\theta,r)$ can be 
expressed as follows:
\begin{equation}  
\bar{\nu}(\theta,r) = \frac{1}{N_c}\sum_{i \in \{N_c\}}\nu_i(\theta,r).
\label{eqn3}
\end{equation}
where $N_c$ is the number of critical contacts in the pile.
The evolution of $\bar{\nu}(\theta,r)$ is displayed in 
Fig.~\ref{fig4}(a) for three values of $\theta$. 
Independantly of the inclination $\theta$, $\bar{\nu}$
decreases rapidly as a function of $r$ as long as $r \lesssim 10 D$. 
Beyond this distance, 
$\bar{\nu}$ remains almost constant and equal to the 
state of the pile $\nu(\theta, V_{pile})$.
Such an evolution of $\bar{\nu}(\theta,r)$ can be approximated
by (see Fig~\ref{fig4}(b)):
\begin{equation}
\label{eqn4}
\bar{\nu}(\theta,r)= \begin{cases}
\nu_{\infty}(\theta)\left(1+A(\theta)r^{\alpha(\theta)}\right)& \text{if $r \leq 10D$},\\
\nu_{\infty}(\theta)& \text{if $r >10D$}.
\end{cases}  
\end{equation}
where $\nu_{\infty}(\theta)$ is the state 
of the pile at $\theta$, and $A(\theta)$ and 
$\alpha(\theta)$ are affine functions of $\theta$.
The distance $L \simeq 10D$ arises here as the characteristic length
of the spatial correlations between critical contacts.   
\\
\indent The spatial correlations can be illustrated when looking at  
two snapshots of the pile for different 
values of the inclination angle $\theta$ (see Fig~\ref{fig5}).
In particular, the system exhibits clusters of critical contacts
where $\nu \ge \nu_c$. These clusters 
can no longer sustain shear load increment and eventually 
lead to local rearrangements.
They can be analysed as ``fluidized'' zones with $\theta$ increasing size.\\
\indent We define $r_c$ the mean size of the clusters of
critical contacts characterized by a state $\nu= \nu_c$.
Equation~(\ref{eqn4}) yields  
\begin{equation}
r_c(\theta)= \left[\frac{1}{A(\theta)}\left(\frac{\nu_c}
{\nu_{\infty}}-1\right)\right]^{1/\alpha(\theta)},
\label{eqn5}
\end{equation}
where 
$\nu_\infty$ is defined as the state of the pile $\nu(\theta,V_{pile})$ 
averaged over all the simulations (see inset graph in Fig~\ref{fig1}).
The evolution of $r_c(\theta)$, displayed in Fig~\ref{fig6}, shows a slow 
increase from around $2D$ to $4D$ as long as $\theta \lesssim 15$ degrees.
For larger $\theta$, the size increases quickly with a power-law divergence
$\propto (\theta_c - \theta)^{-\beta}$. That is reminescent of a percolation-like 
process. The value $\theta=15$ degrees may be related to the dynamical angle of repose of the pile, but this would require further investigation.\\
\indent The mobilization of the granular pile is related to the occurence of ``fluidized'' clusters of critical contacts. The size of these clusters increases with $\theta$, with a power-law divergence as $\theta \rightarrow \theta_c$. This result is in support of a description of the granular system during mobilization as a multi-phase system. The transition from a static equilibrium to a
dynamical flow is controlled by the order parameter $\Phi =\nu/\nu_c$, and the stability threshold would result from a multi-phase instability. Detailed analysis of the phases interaction during transition still 
requires further analysis. \\
%
%
\indent Forces in granular media have been shown to obey a broad distribution 
and to define a strong and a weak contact network. Forces transmitted 
along the strong contact network exceed the mean force in the media and are
responsible for the mechanical strength of the medium, 
with a typical correlation length $\xi \simeq 10D$~\cite{radjai98,mueth98}.
 By contrast, forces transmitted along the weak contact network mainly
 contributes to an average pressure.
In an extended forthcoming paper, we show that critical contacts appear
mostly within the weak contact network. They correspond to lower pressure
as a result of a screening effect by the strong contact network that support the deviatoric load and controls the static stress evolution.
Critical contacts tend to be organized as ``fluidized'' clusters that 
eventually destabilize the pile when their size becomes comparable with 
the correlation length of strong forces $\xi$. The avalanche instability 
appears as a two phase instability triggered by the interaction between 
these correlated clusters and a solid skeleton made  
of the strong contact network.
The influence of material parameters such as the coefficient 
of friction and the degree of configurational disorder in the pile
are the purpose of further work. The extension of
the problem to three dimensions has to be investigated. 

We gratefully acknowledge S. Roux and D. Sornette for useful 
discussions and suggestions about this work. This research 
was supported by the French Ministry of Research and the CNRS 
through the ACI ``Catastrophes Naturelles'' 
and the GdR ``Milieux Divis\'es''.     


\newpage

\begin{figure}[ht]
\caption{\label{fig1}(a) Density $\nu$ of critical contacts in the pile
as a function of the tilt angle $\theta$  for a single run (main curve) and averaged over 15 independant runs (inset curve),   
(b) Mean translational kinetic
energy $E_k$ ($kg.m^2.s^{-2}$) of particles  
as a function of $\theta$ for the same single run, (c) mean envelope of the
maximum values of $\nu$ evaluated over the 15 runs. }
\end{figure} 

\begin{figure}[ht]
\caption{\label{fig2} Proportion $\Delta N/N$ of contacts lost in the packing 
during rearrangements as a function of $\nu$.}
\end{figure} 

\begin{figure}[ht]
\caption{\label{fig3} Probability density function (pdf) of 
local states $\nu$ in the neighbourghood of critical contacts (filled circles) 
and non critical contacts (opaque circles), for a neighbourghood radius  
 $r=6D$.}
\end{figure}

\begin{figure}[ht]
\caption{\label{fig4} Mean state $\bar{\nu}$ in the neighbourhood of a critical contact as a fonction of the size $r$ of the neighbourhood (a) in linear scale for three values of the tilt angle  $\theta = 6^\circ$ (solid line), $\theta=8^\circ$ (dashed line) and $\theta=12^\circ$ (dotted line), and (b) in logarithmic scale for $\theta = 6^\circ$ (solid line) shown with a line with a slope $\alpha =-1.5$ (dot-dashed line).}
\end{figure}

\begin{figure}[ht]
\caption{\label{fig5} Density of critical contacts in the pile at
 $\theta \simeq 5^\circ$ (upper picture) and at $\theta \simeq 16^\circ$ (bottom picture).
Areas where $\nu \ge \nu_c$ are in black, whereas the state $\nu=0$ are in white. }
\end{figure}

\begin{figure}[ht]
\caption{\label{fig6}Evolution of the mean size $r_c$ of 
 clusters of critical contacts (see text) 
as a function of the tilt angle $\theta$ in linear scale
 (main curve) and in logarithmic scale (inset curve), where
 $\theta_c = 19.96$ degrees and the line 
is drawn with a slope $-\beta \simeq 0.5$. }
\end{figure}

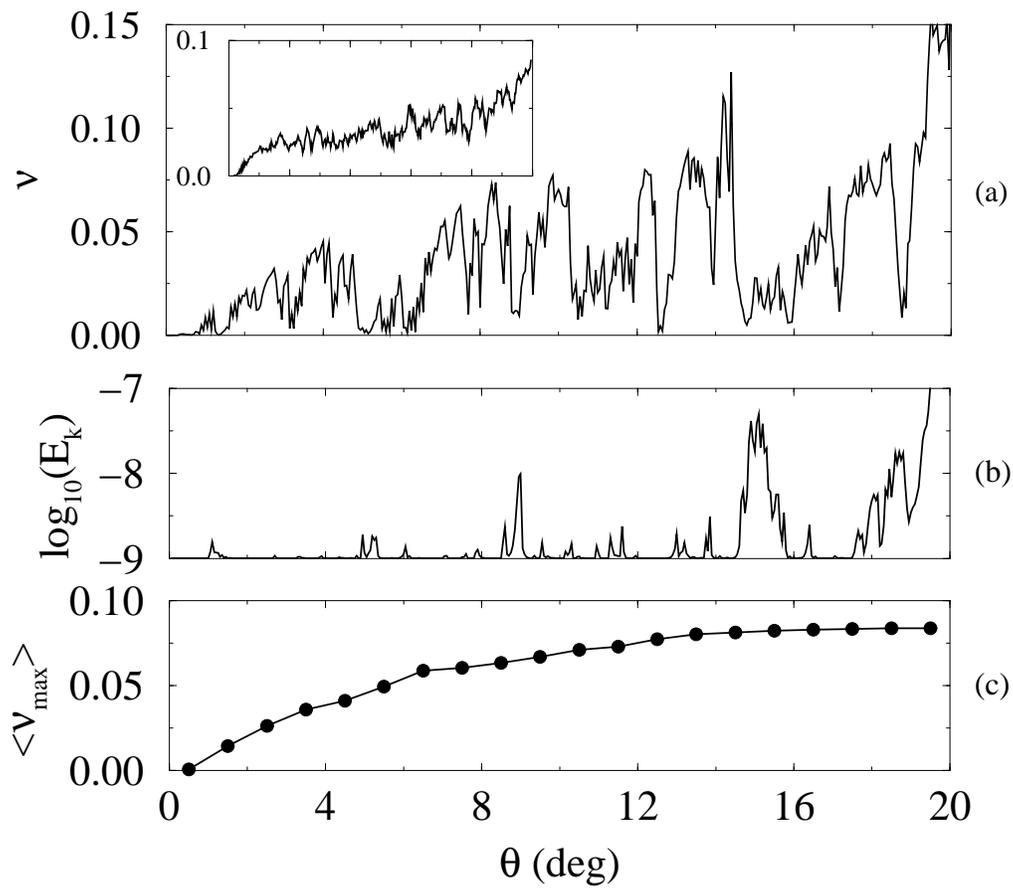
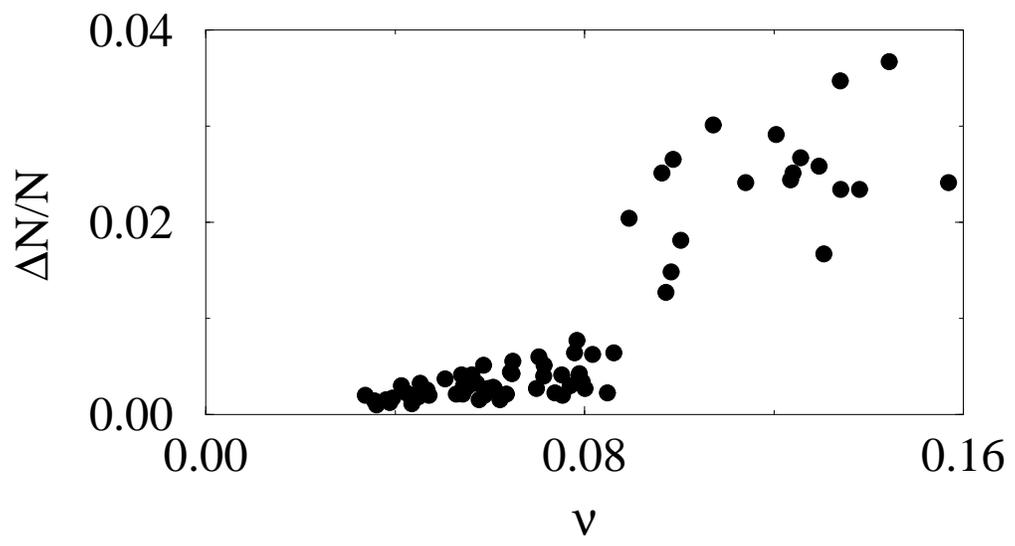
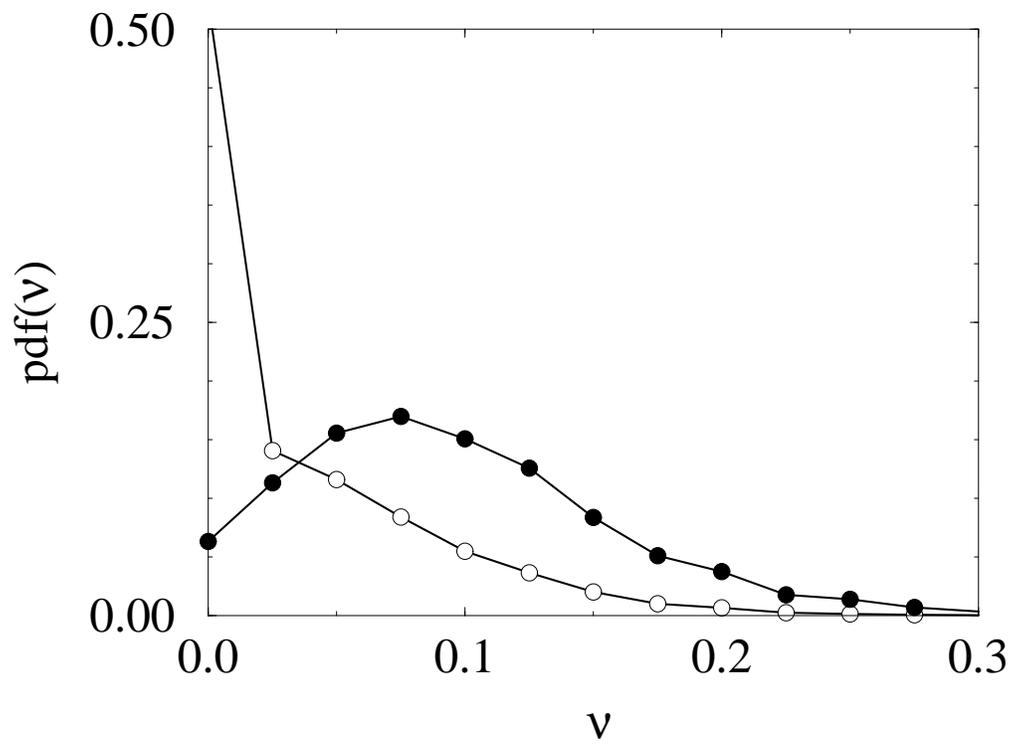
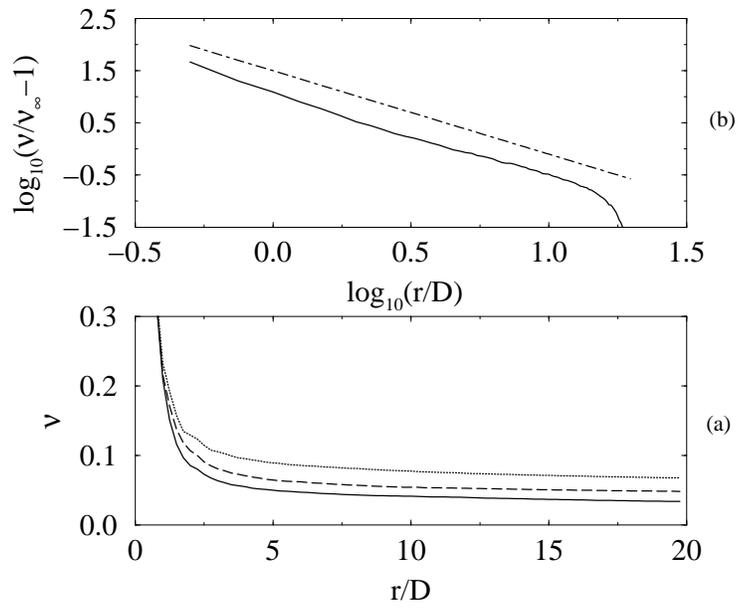
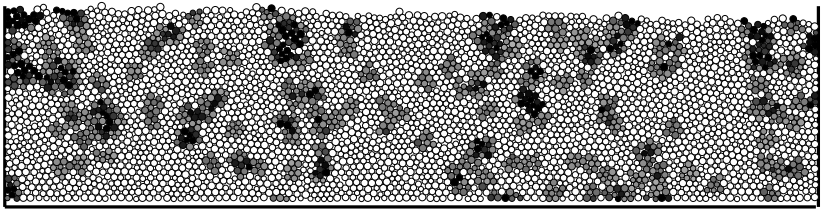
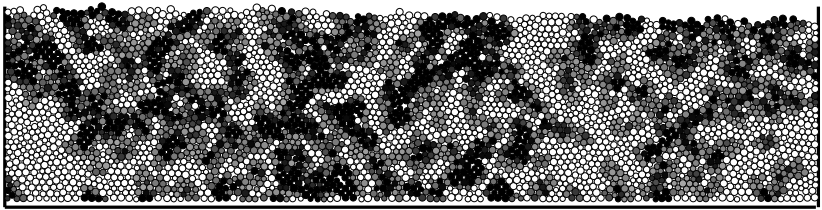
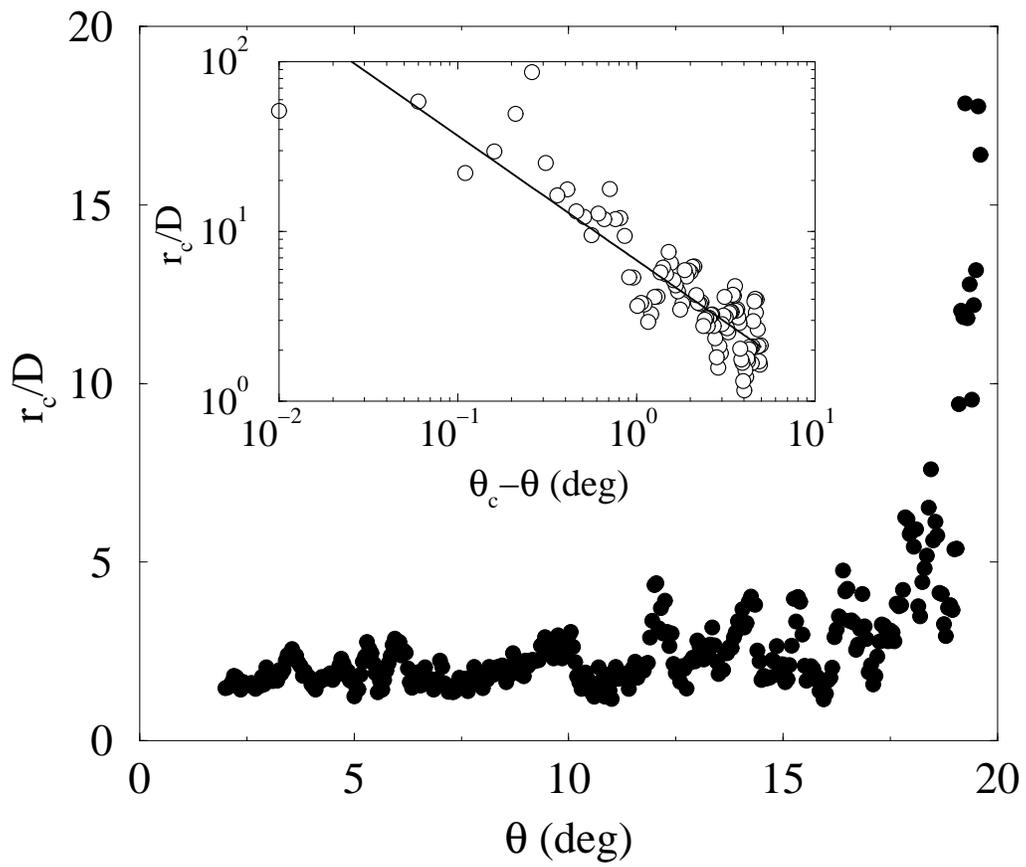
\end{document}